\documentclass [] {aa} 
\usepackage{epsfig}

\begin{document}

\thesaurus{06(08.01.1; 08.03.02; 08.16.2)}

\title{Chemical analysis of 8 recently discovered extra-solar planet host stars\thanks{Based on
    observations collected at the La Silla Observatory, ESO (Chile),
    with the echelle spectrograph {\footnotesize CORALIE} at the 1.2-m
    Euler Swiss telescope}}


\author{N.C.~Santos\inst{1} \and G.~Israelian\inst{2} \and M.~Mayor\inst{1}} 

\institute{Observatoire de Gen\`eve, 51 ch.  des
	   Maillettes, CH--1290 Sauverny, Switzerland \and 
	   Instituto de Astrofisica de Canarias, E--38200 La Laguna, Tenerife, Spain}

\date{Received / Accepted } 
\offprints{Nuno.Santos@obs.unige.ch}

\titlerunning{Chemical analysis of 8 planetary host stars} 
\maketitle


\begin{abstract}  

We present the chemical analysis of a new set of stars known to harbor low
mass companions, namely \object{{\footnotesize HD}\,1237},
\object{{\footnotesize HD}\,52265},
\object{{\footnotesize HD}\,82943}, \object{{\footnotesize HD}\,83443},
\object{{\footnotesize HD}\,169830}, and \object{{\footnotesize HD}\,202206}.
In addition, we have also analyzed
\object{{\footnotesize HD}\,13445} and \object{{\footnotesize HD}\,75289},
already studied elsewhere. The abundances of C and $\alpha$-elements O, S, Si, 
Ca and Ti are presented and discussed in the context of the metallicity 
distribution of the stars with extra-solar planets. We
compare the metallicity distribution of stars with planets with the 
same distribution of field dwarfs. The results further confirm that stars 
with planets are over-abundant in [Fe/H].

\end{abstract}

\section{Introduction}

Before the discovery of the first extra-solar planet (Mayor \& Queloz \cite{May95}), the constraints 
on the planetary formation models were confined to the Solar System example. 
Today more than 40 extra-solar planetary systems are known. These discoveries 
immediately opened new horizons to this field, but also brought a bunch of new questions and problems.
In fact, the extra-solar planets found to date don't have 
much in common with our own solar system. Their physical 
characteristics were completely unexpected. Some of them, like the extreme proximity 
of some of the new planets to their ``mother'' stars are surprising and still defy current 
formation models. A review of the subject can be found in Marcy, Cochran \& Mayor (\cite{Mar99}).

One particular fact became evident soon after the first extra-solar 
planets were discovered. The metallicity of the stars with planets 
proved to be distinctively different from the one found in field single dwarfs, being 
very metal-rich (Gonzalez \cite{Gon97}, \cite{Gon98a}). 
As the number of extra-solar planets is increasing, 
this fact is becoming more and more sharp.

Recent work showed that the 
high metallicity of the stars with giant planets cannot be the result of stellar population 
effects (Gonzalez \cite{Gon99}). Furthermore,
Gonzalez \& Laws (\cite{Gon00}) add more evidence for chemical 
anomalies in stars with planetary companions. Their study suggests that the anomalies
not only involve the [Fe/H] index, but possibly also the ratios [Li/H], [C/H] and 
[N/H]. The observed correlations between the presence of planets and the existence of
chemical anomalies represent the only 
known physical connection between their presence and a stellar 
photospheric parameter, and their study is thus of major importance.

In addition to the discovery of a planet around \object{{\footnotesize HD}\,1237} 
(Naef et al. \cite{Nae00}), the Geneva extra-solar planet search group recently 
announced the discovery of 8 new low-mass companions to solar type stars (all with masses below $\sim$15\,$M_\mathrm{Jup}$\footnote{www.eso.org/outreach/press--rel/pr--2000/pr--13--00.html}). These new planets and brown-dwarfs are included in a volume-limited sample of dwarfs 
(Udry et al. \cite{Udr00}).
In this paper we present and discuss the abundance analysis of the elements C, O, S, Ca, Ti, Fe, 
and Si in the new candidates, \object{{\footnotesize HD}\,1237}, 
\object{{\footnotesize HD}\,52265}, \object{{\footnotesize HD}\,82943}, 
\object{{\footnotesize HD}\,83443}, \object{{\footnotesize HD}\,169830}, and 
\object{{\footnotesize HD}\,202206}. 
Planets have also been discovered during the {\footnotesize CORALIE} survey 
around \object{{\footnotesize HD}\,13445} (Queloz et al. \cite{Que00}) and 
\object{{\footnotesize HD}\,75289} (Udry et al. \cite{Udr00}). 
We also present a spectroscopic analysis of these two stars,
already studied by Flynn \& Morell (\cite{Fly97}), and Gonzalez \& Laws (\cite{Gon00}), 
respectively. The other three recently announced candidates 
(\object{{\footnotesize HD}\,108147}, \object{{\footnotesize HD}\,162020} and 
\object{{\footnotesize HD}\,168746}) will be the subject of a future publication.

\section{Observations and data reduction}

\begin{figure}
\psfig{width=\hsize,file=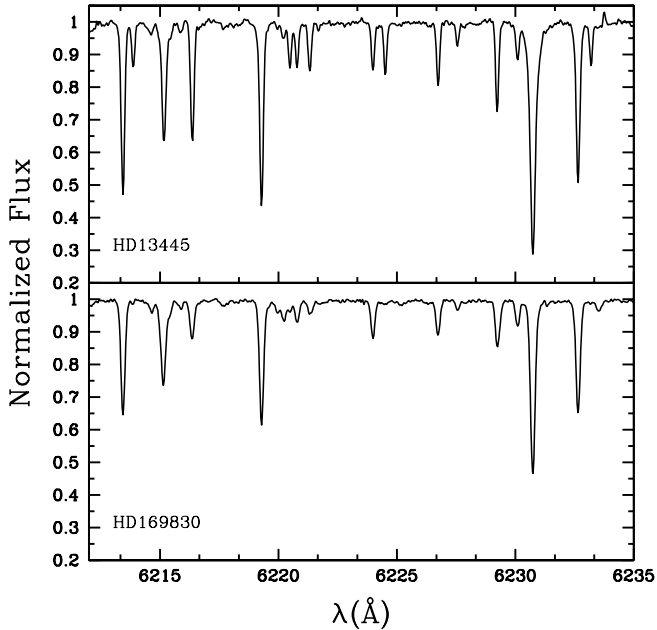}
\caption[]{Two sample {\footnotesize CORALIE} spectra of the K0 dwarf \object{{\footnotesize HD}\,13445} 
(upper panel), and of the F8 dwarf \object{{\footnotesize HD}\,169830} (lower panel).}
\label{fig1}
\end{figure}

The spectra were obtained during three separate runs, between January and April 2000,
using the new 1.2-m Euler Swiss telescope at La Silla (ESO), 
Chile, equipped with the {\footnotesize CORALIE} echelle spectrograph.
The resolving power ($\lambda/\Delta\lambda$) of the spectrograph is about 50\,000, 
and the spectra cover the visible spectrum between 3800 and 6800\,\AA\ without gaps. 
The spectra have S/N ratios between about 150 and 350 (for the faintest objects 
we doubled the exposures). In Fig.~\ref{fig1} we present two sample spectra of 
\object{{\footnotesize HD}\,13445} and \object{{\footnotesize HD}\,169830}.

The reduction of the spectra was carried out using standard tasks in the 
{\tt echelle} package of IRAF\footnote{IRAF is distributed by National Optical Astronomy 
Observatories, operated by the Association of Universities for Research in Astronomy, 
Inc., under contract with the National Science Foundation, U.S.A.}. 
The wavelength calibration was done using the 
spectrum of a Thorium-Argon lamp that was taken in the beginning of the night. 

After wavelength calibrated, the spectra were corrected for the radial-velocity 
Doppler shift using the velocity computed in the context of the planet search programme,
and normalized using the {\footnotesize CONTINUUM} task in IRAF. In this step 
we divided the spectra into 350\,\AA\ wide intervals, that were normalized separately 
using 3rd order spline functions, and added together at the end. A visual inspection 
of the resulting spectra showed that the results were quite satisfactory. 

\section{Analysis}

\subsection{Spectrum Synthesis: Method}

Abundance determination was done using a standard Local Thermodynamic Equilibrium (LTE) 
analysis with a revised version of the line abundance code MOOG
(Sneden \cite{Sne73}), and a grid of Kurucz et al. (\cite{Kur93}) ATLAS9 atmospheres.

Atomic data for iron lines was taken from the list of
Gonzalez \& Laws (\cite{Gon00}). This choice was done so that 
our results would be in the same scale with theirs. Since we could not 
obtain a Solar spectrum, we also used the gf-values listed by these authors. 

The line-lists for other elements were compiled from different authors, and 
semi-empirical gf-values were computed using equivalent widths obtained in the High 
Resolution Solar Atlas (Kurucz et al. \cite{Kur84}), and a solar model with 
$T_\mathrm{eff}$~=~5777\,K, $\log{g}$~=~4.438 and $\xi_t$~=~1\,km\,s$^{-1}$. 
We decided to adopt a value of $\log{\epsilon_{\sun}}(Fe)$~=~7.47 (the same used 
by Gonzalez \& Laws \cite{Gon00}). For other elements, the values were taken 
from Anders \& Grevesse (\cite{And89}).

Measured equivalent widths for the lines in the observed stars were determined by Gaussian fitting using 
the SPLOT task in IRAF. In Tables~\ref{tab1} and \ref{tab2} we present the results, as well as 
the line parameters used ($\log{gf}$ and $\chi_l$).

\begin{table*}
\caption[]{\ion{Fe}{i} and \ion{Fe}{ii} line parameters and measured equivalent widths. }
\begin{tabular}{lccrrrrrrrr}
\hline
\noalign{\smallskip}
$\lambda_{0}$ & $\chi_{l}$ & $\log{gf}$ & \object{{\footnotesize HD}\,1237} & \object{{\footnotesize HD}\,13445} & \object{{\footnotesize HD}\,52265} & \object{{\footnotesize HD}\,75289} & \object{{\footnotesize HD}\,82943} & \object{{\footnotesize HD}\,83443} & \object{{\footnotesize HD}\,169830} & \object{{\footnotesize HD}\,202206} \\
            &          &            &  \multicolumn{8}{c}{Equivalent Widths (m\AA)}  \\  
\hline \\
\ion{Fe}{i}   &  \multicolumn{10}{l}{$\log{\epsilon_{\sun}}~=~7.47$}    \\\\
5044.21 &2.85 &$-$2.04 &101.7 &95.4 &74.7 &78.1 &83.1 &102.9 &66.9 &91.6\\
5247.05 &0.09 &$-$4.93 &104.2 &86.4 &71.3 &73.2 &69.8 &104.3 &53.1 &82.5\\
5322.05 &2.28 &$-$2.86 &80.1  &69.0 &62.1 &65.0 &65.2 &94.0 &51.5 &75.1\\
5806.73 &4.61 &$-$0.90 &70.7  &55.0 &59.9 &61.9 &64.2 &84.7 &50.1 &75.2\\
5852.22 &4.55 &$-$1.18 &58.4  &44.4 &44.6 &49.1 &52.0 &66.2 &35.9 &57.5\\
5855.09 &4.61 &$-$1.52 &33.0  &22.3 &25.2 &28.5 &29.4 &42.8 &20.2 &35.1\\
5856.09 &4.29 &$-$1.56 &48.2  &34.5 &36.9 &41.3 &41.7 &59.1 &29.6 &47.5\\
6027.06 &4.08 &$-$1.09 &79.3  &65.7 &69.3 &73.9 &76.8 &89.0 &64.5 &82.7\\
6056.01 &4.73 &$-$0.40 &94.6  &77.7 &78.8 &83.0 &83.9 &102.5 &71.7 &94.8\\
6079.01 &4.65 &$-$1.02 &67.4  &45.9 &50.3 &55.2 &55.9 &73.0 &42.9 &67.0\\
6089.57 &5.02 &$-$0.86 &48.7  &36.8 &34.9 &41.2 &41.8 &61.0 &31.5 &51.3\\
6151.62 &2.18 &$-$3.29 &70.2  &59.7 &46.4 &50.0 &55.0 &74.1 &37.0 &64.1\\
6157.74 &4.07 &$-$1.25 &80.4 &62.1 &69.7 &69.2 &74.5 &90.5 &59.9 &84.0\\
6159.38 &4.61 &$-$1.87 &16.9 &12.1 &13.4 &14.5 &17.5 &28.2 &10.6 &24.0\\
6165.37 &4.14 &$-$1.47 &57.1 &48.0 &47.0 &51.5 &53.6 &66.1 &40.1 &59.7\\
6180.21 &2.73 &$-$2.61 &81.0 &65.0 &60.9 &67.1 &66.5 &92.9 &50.4 &78.2\\
6187.99 &3.94 &$-$1.61 & --  & --  &49.1 & --  &--   &--   &--   &--\\
6200.32 &2.61 &$-$2.44 &99.0 &85.2 &74.5 &77.2 &82.1 &106.1 &64.6 &95.6\\
6226.74 &3.88 &$-$2.03 &42.6 &31.9 &34.5 &32.0 &37.7 &54.4 &24.3 &47.4\\
6229.23 &2.84 &$-$2.82 &57.3 &44.3 &40.5 &44.8 &47.3 &67.4 &33.8 &59.6\\
6240.65 &2.22 &$-$3.32 &73.8 &58.6 &47.4 &51.3 &55.5 &75.2 &36.8 &65.6\\
6265.14 &2.18 &$-$2.57 &119.3 &102.6 &85.5 &89.7 &91.0 &120.1 &79.7 &105.5\\
6270.22 &2.86 &$-$2.57 &72.0 &58.5 &53.4 &59.5 &59.1 &77.4 &49.1 &69.8\\
6380.75 &4.19 &$-$1.32 &66.3 &52.5 &54.1 &59.1 &61.7 &85.0 &47.8 &68.1\\
6392.54 &2.28 &$-$4.01 & --&28.0 &14.9 &16.3 &18.5 &41.2 &9.3 &28.6\\
6498.95 &0.96 &$-$4.62 &68.3 &64.9 &42.4 &44.2 &51.6 &78.7 &29.0 &64.3\\
6591.33 &4.59 &$-$1.98 &15.7 &11.8 &13.4 &11.4 &16.5 &25.1 &11.8 &22.6\\
6608.04 &2.28 &$-$4.00 &30.1 &23.4 &14.7 &17.5 &22.2 &46.4 &11.2 &33.7\\
6627.56 &4.55 &$-$1.44 &43.1 &30.9 &34.5 &36.5 &37.0 &51.8 &27.8 &44.6\\
6646.93 &2.61 &$-$3.85 &18.6 &15.0 &14.6 &11.1 &14.1 &32.4 &7.0 &22.0\\
6653.91 &4.15 &$-$2.41 & -- &12.2 &-- &-- & --&-- &10.5 &20.0\\
6703.58 &2.76 &$-$3.01 &56.5 &47.5 &34.1 &39.8 &42.6 &67.4 &27.6 &48.7\\
6710.31 &1.48 &$-$4.80 &29.6 &27.9 &13.3 &11.6 & & &9.0 &29.4\\
6725.36 &4.10 &$-$2.18 &26.8 &17.7 &18.6 &22.8 &23.9 &37.8 &15.0 &29.2\\
6726.67 &4.61 &$-$1.04 &63.1 &50.8 &51.2 &54.3 &55.2 &70.8 &42.8 &66.8\\
6733.15 &4.64 &$-$1.45 &40.6 &28.6 &30.1 &29.9 &36.6 &49.0 &22.8 &40.4\\
6745.11 &4.58 &$-$2.06 &14.1 &8.7 &10.6 &12.2 &14.3 &25.5 &-- &--\\
6750.15 &2.42 &$-$2.62 &96.5 &87.2 &74.4 &74.4 &79.8 & --&66.2 &89.9\\
6752.72 &4.64 &$-$1.20 &54.3 &39.2 &37.8 &40.5 &46.5 &71.4 &32.7 &54.2\\
6786.87 &4.19 &$-$1.95 &40.8 &28.7 &28.8 &35.1 &36.3 &51.7 &21.5 &44.5\\\\
\ion{Fe}{ii} &  \multicolumn{10}{l}{$\log{\epsilon_{\sun}}~=~7.47$}    \\\\
5234.63 &3.22 &$-$2.20 &88.5 &60.8 &106.9 &106.9 &99.4 &94.2 &114.9 &94.2\\
6084.11 &3.20 &$-$3.75 &17.7 &8.6 &35.5 &37.5 &30.9 &28.9 &39.8 &28.9\\
6149.25 &3.89 &$-$2.70 &33.2 &16.2 &57.7 &55.5 &51.0 &41.0 &61.0 &41.0\\
6247.56 &3.89 &$-$2.30 &55.7 &25.1 &80.4 &83.9 &72.5 &52.1 &88.5 &52.1\\
6369.47 &2.89 &$-$4.11 &19.0 &6.5 &31.8 &34.2 &30.7 &22.4 &36.9 &22.4\\
6416.93 &3.89 &$-$2.60 &43.6 &24.4 &56.5 &55.8 &53.2 &49.1 &62.5 &49.1\\
6432.68 &2.89 &$-$3.29 &39.1 &18.8 &58.8 &63.6 &53.1 &46.6 &66.0 &46.6\\
\\
\noalign{\smallskip}
\hline
\end{tabular}
\label{tab1}
\end{table*}

\begin{table*}
\caption[]{Line parameters and measured equivalent widths for C, O, S, Ca, Ti and Si lines. }
\begin{tabular}{lccrrrrrrrr}
\hline
\noalign{\smallskip}
$\lambda_{0}$ & $\chi_{l}$ & $\log{gf}$ & \object{{\footnotesize HD}\,1237} & \object{{\footnotesize HD}\,13445} & \object{{\footnotesize HD}\,52265} & \object{{\footnotesize HD}\,75289} & \object{{\footnotesize HD}\,82943} & \object{{\footnotesize HD}\,83443} & \object{{\footnotesize HD}\,169830} & \object{{\footnotesize HD}\,202206} \\
            &          &            &  \multicolumn{8}{c}{Equivalent Widths (m\AA)}  \\    
\hline \\
\ion{C}{i} &  \multicolumn{10}{l}{$\log{\epsilon_{\sun}}~=~8.56$}    \\\\
5380.34 & 7.68 & $-$1.71 & 17.0 &  7.8 & 39.4 & 39.0 & 36.3 & 26.1 & 45.1 & 22.5\\
6587.61 & 8.53 & $-$1.08 &  9.9 & --   & 29.6 & 27.8 & 24.9 & --   & 36.2 & 16.7\\\\
\ion{O}{i} &  \multicolumn{10}{l}{$\log{\epsilon_{\sun}}~=~8.93$}    \\\\
6300.30 & 0.00 & $-$9.84 & --   & --   &  5.1 & 3.2  &  5.2 &  9.7 & --   & --  \\\\
\ion{S}{i} &  \multicolumn{10}{l}{$\log{\epsilon_{\sun}}~=~7.21$}    \\\\
6046.03 & 7.87 & $-$0.23 & 15.1 & --   & 27.3 & 28.3 & 29.0 & 25.4 & 31.6 & 18.0\\
6052.67 & 7.87 & $-$0.44 & 8.8  & --   & 24.1 & 21.2 & 20.1 & 18.7 & 26.0 & 14.3\\\\
\ion{Ca}{i} & \multicolumn{10}{l}{$\log{\epsilon_{\sun}}~=~6.36$}    \\\\
5867.57 & 2.93 & $-$1.56 & 40.8 & 36.3 & 25.5 & 31.6 & 29.4 & 49.2 & 17.6 & 40.4\\
6166.44 & 2.52 & $-$1.10 & 90.3 & 91.1 & 72.1 & 75.4 & 76.9 &101.1 & 61.0 & 89.2\\
6169.05 & 2.52 & $-$0.68 &137.9 &128.1 & 97.5 &102.0 &101.1 &138.7 & 86.9 &116.4\\
6455.62 & 2.52 & $-$1.34 & 77.4 & 69.4 & 61.9 & 62.2 & 64.9 & 88.5 & 52.8 & 75.0\\
6471.66 & 2.53 & $-$0.80 &123.0 &118.5 &112.7 &101.5 &108.4 &127.9 & 87.7 &110.0\\
6499.65 & 2.52 & $-$0.90 &114.9 &106.2 & 88.8 & 93.1 & 95.4 &118.1 & 82.7 &108.9\\
6508.81 & 2.53 & $-$2.33 & 32.6 & 16.3 &  7.4 & 14.9 & 14.4 & 26.8 &  6.2 & 24.5\\\\
\ion{Ti}{i} & \multicolumn{10}{l}{$\log{\epsilon_{\sun}}~=~4.99$}    \\\\
5087.06 & 1.43 & $-$0.88 & 53.1 & 69.8 & 25.2 & 31.3 & 33.1 & 78.1 & 18.0 & 48.0\\
5113.45 & 1.44 & $-$0.91 & 48.2 & 65.0 & 24.7 & 27.5 & 30.5 & 60.1 & 15.3 & 44.1\\
5300.01 & 1.05 & $-$1.47 & 28.4 & 28.1 & --   & --   & --   & 52.0 & 17.5 & --  \\
5426.25 & 0.02 & $-$3.05 & 18.1 & 26.2 & --   & --   & 8.2  & 33.2 &  3.0 & 12.1\\
5866.46 & 1.07 & $-$0.92 & 72.3 & 83.3 & 46.3 & 48.8 & 51.3 & 91.4 & 30.1 & 65.1\\
5965.84 & 1.88 & $-$0.38 & 53.2 & 55.1 & 36.0 & 30.7 & 45.4 & 80.3 & 21.7 & 46.3\\
6126.22 & 1.07 & $-$1.41 & 43.0 & 48.0 & 18.2 & 19.5 & 24.0 & 55.0 & 13.0 & 43.0\\
6261.11 & 1.43 & $-$0.46 & 74.1 & 81.4 & 46.2 & 49.2 & 53.1 & 94.8 & 32.9 & 70.2\\\\
\ion{Ti}{ii}& \multicolumn{10}{l}{$\log{\epsilon_{\sun}}~=~4.99$}    \\\\
 4589.96 & 1.23 & $-$1.61 & 89.2 & 74.6 &101.4 &105.3 & 98.1 &101.3 &106.9 & 94.8\\
 5336.78 & 1.58 & $-$1.61 & 73.2 & 60.1 & 88.0 & 89.5 & 81.8 & 85.0 & 94.1 & 77.0\\
 5418.77 & 1.58 & $-$2.07 & 52.1 & 37.0 & 61.4 & 64.1 & 60.2 & 61.2 & 67.1 & 57.1\\\\
\ion{Si}{i} & \multicolumn{10}{l}{$\log{\epsilon_{\sun}}~=~7.55$}    \\\\
 5665.56 & 4.92 & $-$1.98 & 51.0 & 34.3 & 50.6 & 50.7 & 53.1 & --   & 40.3 & 61.6\\
 5690.43 & 4.93 & $-$1.82 & 59.5 & 41.9 & 56.9 & 61.4 & 62.1 & 72.7 & 52.5 & 65.6\\
 5793.07 & 4.93 & $-$1.96 & 52.2 & 36.3 & 55.7 & 57.6 & 59.0 & 73.2 & 50.7 & 61.7\\ 
 5948.54 & 5.08 & $-$1.08 &101.1 & 87.8 &105.7 & 97.5 &118.5 &131.7 & 86.7 &109.0\\ 
 6091.91 & 5.87 & $-$1.36 & 48.1 & 26.6 & 53.1 & 55.0 & 58.9 & 89.2 & 40.8 & --  \\
 6125.01 & 5.61 & $-$1.51 & 39.9 & 26.9 & 45.5 & 45.2 & 50.7 & 57.4 & 37.2 & 53.4\\
 6142.47 & 5.62 & $-$1.48 & 35.3 & 25.7 & 46.0 & 49.5 & 52.9 & 61.8 & 38.4 & 51.9\\
 6155.15 & 5.62 & $-$0.72 & 97.2 & 66.2 &105.6 &107.2 &118.1 &131.3 & 90.5 &114.6\\
 6237.34 & 5.61 & $-$1.06 & 78.6 & 48.2 & 85.9 & 88.6 & 90.2 &116.0 & 71.4 & 94.0\\
 6583.69 & 5.95 & $-$1.65 & 18.9 & 14.7 & 34.2 & 37.5 & 34.3 & 44.6 & 21.4 & 40.7\\
 6721.86 & 5.86 & $-$1.14 & 53.7 & 34.1 & 60.6 & 63.9 & 65.7 & 85.2 & 50.2 & 73.5\\
\\
\noalign{\smallskip}
\hline
\end{tabular}
\label{tab2}
\end{table*}

\subsection{Stellar parameters and Abundances}

\begin{figure}
\psfig{width=\hsize,file=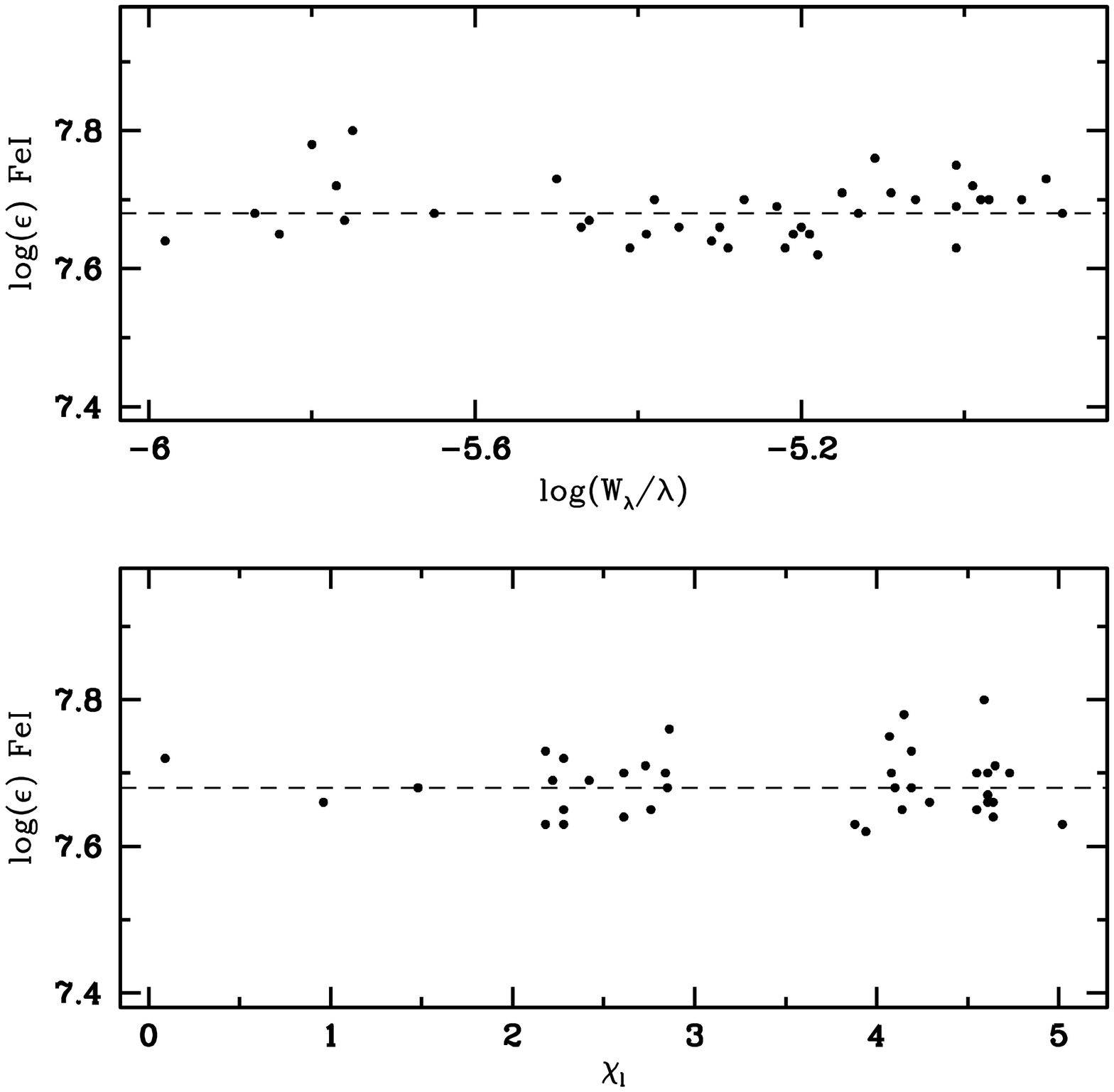}
\caption[]{\ion{Fe}{i} abundances for \object{{\footnotesize HD}\,169830} computed using the
atmospheric parameters from Table~\ref{tab3}. The dashed lines represent the
fit to the points. {\it Upper panel}: abundance vs. reduced equivalent width; {\it lower panel}:
abundance vs. lower excitation potential.}
\label{fig2}
\end{figure}

Stellar atmospheric parameters were computed using the standard technique based on the Fe ionization balance. We first adopted initial atmospheric parameters 
computed from {\it ubvy}-photometry of Hauck \& Mermilliod (\cite{Hau97}) using the 
calibrations of Olsen (\cite{Ols84}) for $T_\mathrm{eff}$ and $\log{g}$, Schuster \& 
Nissen (\cite{Sch89}) for [Fe/H] and Edvardsson et al. (\cite{Edv93}) for $\xi_t$. Then, 
using the set of \ion{Fe}{i} and \ion{Fe}{ii} lines presented in Table~\ref{tab1}, 
we iterated until the correlation coefficients between $\log{\epsilon}$(\ion{Fe}{i}) and $\chi_l$, and 
between $\log{\epsilon}$(\ion{Fe}{i}) and  $\log{(\mathrm{W}_\lambda/\lambda)}$ were zero 
(Fig.~\ref{fig2}). The abundances derived from the \ion{Fe}{ii} lines were forced to be equal
to those obtained from \ion{Fe}{i}. 
The final atmospheric parameters, as well as the resulting iron abundances, 
are summarized in Table~\ref{tab3}. 

Errors in the parameters were estimated in the same way as in Gonzalez \& Vanture (\cite{Gon98b}). 
The values were rounded to 25~K in $T_{\mathrm{eff}}$, 0.05~dex in $\log{g}$, and 0.05~km\,s$^{-1}$ 
in $\xi_t$.

Uncertainties in the abundances of all elements were then determined adding the errors 
due to the sensitivities of the resulting abundances to changes of the atmospheric 
parameters (see Table~\ref{tab4} for two examples), and the dispersion of
the abundances for the individual lines of each element. 
For elements with only one line measured, the errors only take into account the
sensitivities to the atmospheric parameters. The dependence on 
$\xi_t$ is always very small, and has no important implications in the final
errors. The final abundance determinations and errors for C, O, S, Ca, 
\ion{Ti}{i}, \ion{Ti}{ii}, and Si are presented in Table~\ref{tab5}.

\begin{table}
\caption[]{Stellar parameters determined from the Fe lines. }
\begin{tabular}{llccc}
\hline
\noalign{\smallskip}
HD  & $T_\mathrm{eff}$ & $\log{g}$       & $\xi_t$ & [Fe/H] \\
number       & (K)              &  (cm\,s$^{-2}$) & (km\,s$^{-1}$) &  \\
\hline \\
{1237}   &  5540$\pm$75  & 4.70$\pm$0.20  &  1.47$\pm$0.1 &  $+$0.10$\pm$0.08 \\
{13445}  &  5180$\pm$75  & 4.75$\pm$0.25  &  0.79$\pm$0.1 &  $-$0.21$\pm$0.07 \\
{52265}  &  6060$\pm$50  & 4.29$\pm$0.25  &  1.29$\pm$0.1 &  $+$0.21$\pm$0.06 \\
{75289}  &  6140$\pm$50  & 4.47$\pm$0.20  &  1.47$\pm$0.1 &  $+$0.28$\pm$0.07 \\
{82943}  &  6010$\pm$50  & 4.62$\pm$0.20  &  1.08$\pm$0.1 &  $+$0.32$\pm$0.06 \\
{83443}  &  5460$\pm$100 & 4.55$\pm$0.25  &  1.05$\pm$0.1 &  $+$0.38$\pm$0.11 \\
{169830} &  6300$\pm$50  & 4.11$\pm$0.25  &  1.37$\pm$0.1 &  $+$0.21$\pm$0.05 \\
{202206} &  5750$\pm$75  & 4.80$\pm$0.20  &  0.96$\pm$0.1 &  $+$0.36$\pm$0.08 \\
\noalign{\smallskip}
\hline
\end{tabular}
\label{tab3}
\end{table}

\begin{table}
\caption[]{Sensitivities of [X/H] values due to changes in $T_{\mathrm{eff}}$ of $+$100\,k, and 
$\log{g}$ of $+$0.2\,dex for \object{{\footnotesize HD}\,13445} and \object{{\footnotesize HD}\,169830}.}
\begin{tabular}{lrr}
\hline
\noalign{\smallskip}
Star/Element & $\Delta\,T_\mathrm{eff}$~=~+100\,k & $\Delta\,\log{g}$~=~+0.2\,dex \\
\hline \\

\object{{\footnotesize HD}\,13445}  & & \\
 C             & $-$0.08 & +0.08    \\
 Ca            & +0.09   & $-$0.06  \\
 \ion{Ti}{i}   & +0.12   & $-$0.03  \\
 \ion{Ti}{ii}  & 0.00    & +0.06    \\
 Si            & $-$0.02 & +0.02    \\
 Fe            & +0.05   & +0.01    \\
 \\
\hline \\
\object{{\footnotesize HD}\,169830}  & & \\
 C             & $-$0.06 & +0.07    \\
 S             & $-$0.04 & +0.04    \\
 Ca            & +0.06   & $-$0.02  \\
 \ion{Ti}{i}   & +0.08   & $-$0.01  \\
 \ion{Ti}{ii}  & +0.02   & +0.07    \\
 Si            & +0.04   & $-$0.01  \\
 Fe            & +0.07   & 0.00     \\
\\
\noalign{\smallskip}
\hline
\end{tabular}
\label{tab4}
\end{table}

\begin{table*}
\caption[]{Final [X/H] values. }
\begin{tabular}{lcccccccc}
\hline
\noalign{\smallskip}
ele.  & \object{{\footnotesize HD}\,1237} & \object{{\footnotesize HD}\,13445} & \object{{\footnotesize HD}\,52265} & \object{{\footnotesize HD}\,75289} & \object{{\footnotesize HD}\,82943} & \object{{\footnotesize HD}\,83443} & \object{{\footnotesize HD}\,169830} & \object{{\footnotesize HD}\,202206} \\
\hline \\
 C             & $+$0.01$\pm$0.14 & $+$0.08$\pm$0.11 & $+$0.18$\pm$0.10 & $+$0.15$\pm$0.13 & $+$0.22$\pm$0.13 & $+$0.40$\pm$0.11 & $+$0.10$\pm$0.09 & $+$0.17$\pm$0.10 \\
 O             &     --            &   --             & $+$0.07$\pm$0.09 & $-$0.02$\pm$0.09 & $+$0.25$\pm$0.09 & $+$0.50$\pm$0.09 &   --    &   --    \\
 S             & $+$0.15$\pm$0.10 &   --             & $+$0.13$\pm$0.10 & $+$0.09$\pm$0.06 & $+$0.20$\pm$0.05 & $+$0.46$\pm$0.06 & $+$0.04$\pm$0.07 & $+$0.15$\pm$0.10 \\
 Ca            & $+$0.10$\pm$0.13 & $-$0.23$\pm$0.11 & $+$0.19$\pm$0.19 & $+$0.24$\pm$0.06 & $+$0.24$\pm$0.10 & $+$0.24$\pm$0.10 & $+$0.15$\pm$0.12 & $+$0.25$\pm$0.08 \\
 \ion{Ti}{i}   & $+$0.11$\pm$0.12 & $-$0.01$\pm$0.25 & $+$0.22$\pm$0.08 & $+$0.31$\pm$0.08 & $+$0.34$\pm$0.08 & $+$0.49$\pm$0.18 & $+$0.22$\pm$0.11 & $+$0.34$\pm$0.10 \\
 \ion{Ti}{ii}  & $+$0.02$\pm$0.09 & $-$0.16$\pm$0.07 & $+$0.23$\pm$0.13 & $+$0.29$\pm$0.11 & $+$0.38$\pm$0.12 & $+$0.43$\pm$0.13 & $+$0.28$\pm$0.14 & $+$0.41$\pm$0.12 \\
 Si            & $+$0.13$\pm$0.07 & $-$0.09$\pm$0.09 & $+$0.32$\pm$0.08 & $+$0.33$\pm$0.10 & $+$0.39$\pm$0.06 & $+$0.57$\pm$0.12 & $+$0.23$\pm$0.06 & $+$0.40$\pm$0.09 \\
\\
\noalign{\smallskip}
\hline
\end{tabular}
\label{tab5}
\end{table*}

\subsection{Comparison with former results}

Previous spectroscopic analysis have been reported by different authors 
only for three stars from the present sample.

A value of [Fe/H]~=~$-0.24$ was determined by Flynn \& Morell (\cite{Fly97}) for \object{{\footnotesize HD}\,13445}. Our value of $-$0.21, as well as all the atmospheric 
parameters are in good agreement with those proposed by these authors. The small differences can be understood if we note that
these authors used different models of atmospheres and a small number of Fe lines.

The comparison of our atmospheric parameters with those of 
Gonzalez \& Laws (\cite{Gon00}) for \object{{\footnotesize HD}\,75289} shows that they 
are remarkably similar. This similarity comes probably from the fact that we used the same line-list,
models of atmospheres and the spectrum synthesis tool (MOOG). 
It is also apparent that the use of a different spectrograph/configuration
does not have a significant influence on the final results (e.g. [Fe/H] and
other abundance ratios).
The difference is large only for Ca (more than 0.05 dex). In this paper we add the results 
for oxygen, not determined by these authors. 

\object{{\footnotesize HD}\,169830} was included in the large survey of Edvardsson et al. (\cite{Edv93}). 
They obtained $T_\mathrm{eff}$~=~6382\,K, $\log{g}$~=~4.15 and [Fe/H]~=~$+$0.13. These 
values are perfectly compatible with our results, even
though they used a rather different iron line list and a different model atmosphere.  
The value for [Fe/H] is slightly lower than ours, which may be partially explained 
by the fact that they adopted the meteoritic value of $\log{\epsilon_{\sun}}$(Fe) of 7.51.

Gustafsson et al. (\cite{Gus99}) provided carbon abundance for this star using the 
[\ion{C}{i}] line at 8727.14\,\AA. Taking the 
stellar parameters determined by Edvardsson et al. (\cite{Edv93}), they find [C/H]~=~+0.19, 
0.09 dex higher than our estimate.
Below we will discuss carbon abundances in more detail.

\section{Discussion}

\subsection{Metallicity-giant planet connection}

As we can see from tables~\ref{tab3} and \ref{tab5}, 
the planetary host star candidates studied in this work have abundances above 
Solar for almost all elements, the only exception being \object{{\footnotesize HD}\,13445}. 

It is interesting to notice that 3 of the objects are Super Metal Rich (SMR) candidates (defined has
having [Fe/H]$>$0.2 with 95\% confidence, Taylor \cite{Tay96}), namely \object{{\footnotesize HD}\,82943}, \object{{\footnotesize HD}\,83443} and \object{{\footnotesize HD}\,202206}. \object{{\footnotesize HD}\,83443} has the planet in the closest orbit detected to date 
(only 0.028 A.U.), and one of the least massive companions (1.17 times the mass of Saturn). 
This seems to confirm that even the lowest mass companions orbit metal-rich stars. 

\begin{figure}
\psfig{width=\hsize,file=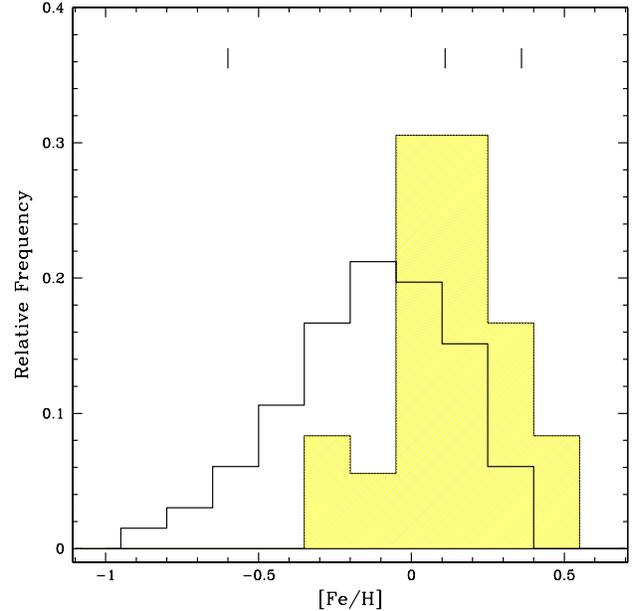}
\caption[]{Metallicity distribution of stars with planets (shaded histogram) 
compared with the same distribution for field G and K dwarfs 
(Favata et al. \cite{Fav97}). The vertical lines represent stars with brown 
dwarf candidate companions.}
\label{fig3}
\end{figure}

In Fig.~\ref{fig3} we compare the distribution of the [Fe/H] values for stars with planets
with the [Fe/H] distribution of a volume-limited sample of field stars (here we used the 
original distribution of Favata et al. (\cite{Fav97}), not corrected for scale galactic height effects). 
This kind of plot was first done by Gonzalez et al. (\cite{Gon98a}) and later
revisited by Butler et al. (\cite{But00}). However, the addition of 6 new planets and
2 brown dwarfs clearly deserves a revision.

The values of [Fe/H] for stars with planets were taken from Table~\ref{tab3} and, 
for the stars not included in this paper, 
from different authors and compiled in Table~4 of Butler et al. (\cite{But00}). 
We did not include the known Brown Dwarf candidates (\object{{\footnotesize HD}\,114762}, \object{{\footnotesize HD162020}\,}, and \object{{\footnotesize HD}\,202206}) in the histogram, but their position in 
the diagram is represented by the vertical lines.
For \object{{\footnotesize HD}\,108147}, \object{{\footnotesize HD}\,162020}, and 
\object{{\footnotesize HD}\,168746} we found no spectroscopic values of [Fe/H] in the 
literature, and we thus determined [Fe/H] from the {\it ubvy}-photometry of Hauck 
\& Mermilliod (\cite{Hau97}), and using the calibration of Schuster \& Nissen (\cite{Sch89}); we obtained [Fe/H]~=~$-$0.02, 0.11, and $-$0.09, respectively. For \object{{\footnotesize HD}\,114762} we took 
the spectroscopic value of [Fe/H]~=~$-$0.60 obtained by Gonzalez (\cite{Gon98a}).

As we can see from Fig.~\ref{fig3}, the addition of the 6 new planets to the histogram further supports the former ideas that stars with planets are particularly metal rich when compared with 
field solar-type dwarfs. This result cannot be related to a selection bias, since the most important
planet search programmes make use of volume limited samples of stars (Udry et al. \cite{Udr00}; Marcy et al. \cite{Mar00}). The only exception is \object{{\footnotesize BD}$-$10\,3166} (Butler et al. \cite{But00}), chosen for its high metallicity. The simple fact that of the 6 new planets announced by
the Geneva group, 5 were discovered around metal-rich dwarfs, clearly shows that the
trend is certainly real.

The position of the three Brown Dwarf candidates (\object{{\footnotesize HD}\,114762}, \object{{\footnotesize HD162020}\,}, and \object{{\footnotesize HD}\,202206}), is intriguing and interesting. Two of the candidates have metallicities that place them perfectly inside the ``planetary'' metallicity 
distribution. This is particularly evident for \object{{\footnotesize HD}\,202206} for which we find [Fe/H]~=~+0.36. 
If the metallicity of the stars with planets is in fact related to their formation process, this fact suggests that the frontier between brown dwarfs and extra-solar giant planets is
not very well defined, and there may be some overlapping.  
But this can also probably be explained if we consider that \object{{\footnotesize HD}\,202206} is in
the metal-rich tail of the field distribution. The inclusion of more brown dwarfs into this plot is 
essential to better clarify this situation.

It is important to refer, however, that we are comparing two distributions 
whose chemical analysis were performed using two different methods. In the present 
article, effective temperatures were derived from Fe lines formed in LTE,
while Favata et al. (\cite{Fav97}) derived them from colours. We do not think that 
this may account for the observed differences in the [Fe/H] distributions. However, we cannot exclude that the use of an uniform analysis may lead to 
somewhat different results.

\subsection{Elements other than Fe}

\begin{figure}
\psfig{width=\hsize,file=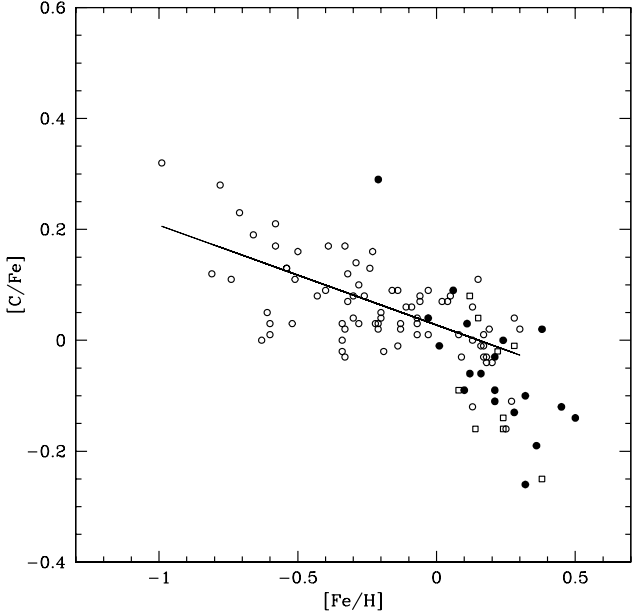}
\caption[]{Plot of [C/H] vs. [Fe/H] for stars with planets (filled circles) and for field stars,
included in the studies of Gustafsson et al. (\cite{Gus99}) and Tomkin et al. (\cite{Tom97}), the open circles and open squares, respectively. The line represents the best linear fit to the 
distribution of Gustafsson et al.}
\label{figc}
\end{figure}

Gonzalez \& Laws (\cite{Gon00}) suggested that stars with planetary companions might be
carbon deficient when compared with field dwarfs.
In Fig.~\ref{figc} we compare the distribution of [C/Fe] for stars with planets (filled circles) to 
the results obtained in the survey of 80 late-F and early-G dwarfs of Gustafsson et al. (\cite{Gus99}), and of the work of Tomkin et al. (\cite{Tom97}). 

Although we have the impression that stars with planets are located below 
the trend for field stars, we believe that statistically we cannot make any serious conclusion. If 
we take only the distribution of Gustafsson et al., we would have the impression that stars with 
planets are positioned below the main trend. However, the addition of the study of Tomkin et al. adds a few points to the region where stars with planets are located.

It is important to mention that all ``three''
studies were done using different sets of lines. Gustafsson et al. used the 8727\AA\ line, while 
the study of Tomkin et al. made use of 6 different carbon lines. Furthermore, the values for the stars with planets are a compilation of the results of this paper and of Gonzalez \& Laws (\cite{Gon00}), Gonzalez (\cite{Gon98b}), and Sadakane et al. (\cite{Sad99}). 
The slight trend that we see, favoring the position of stars with planets in the low part of the 
mean trend line may thus be the result of systematics in the determination of carbon abundances,
connected with the use of different lines and maybe of different atmosphere models (MARCS vs. ATLAS). 
Given that all stars with planets are positioned at the right limit of the plot 
([Fe/H] rich stars), we cannot completely exclude the existence of a change in the slope of 
the relation [C/Fe] vs. [Fe/H] for very metal-rich dwarfs, connected or not with the presence 
of planets.

Given all these points, we prefer to be cautious concerning this result. The resolution of this 
problem may need a consistent and uniform study of carbon abundances in metal rich dwarfs, 
using the same set of lines and atmosphere models.
 
\begin{figure}
\psfig{width=\hsize,file=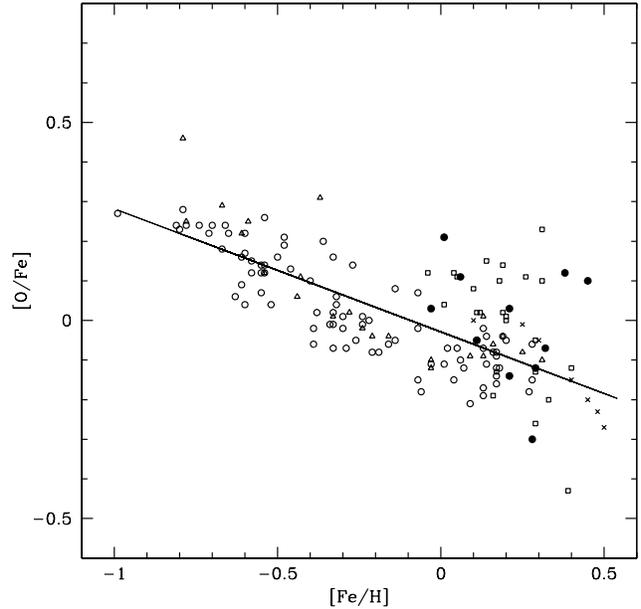}
\caption[]{Plot of [O/Fe] vs. [Fe/H] for stars with planets (filled circles) when compared with the study of field dwarfs of Feltzing \& Gustafsson (\cite{Fel98}), open squares, Edvardsson et al. (\cite{Edv93}), open circles, Castro et al. (\cite{Cas97}), crosses, and Nissen \& Edvardsson (\cite{Nis92}), open triangles. Values for stars with planets were taken from this paper and from Gonzalez \& Vanture (\cite{Gon98b}), Gonzalez (\cite{Gon98a}), and Sadakane et al. (\cite{Sad99}). The best linear fit to the field star data is
also presented.}
\label{figo}
\end{figure}
 
Oxygen and other $\alpha$-elements are produced in massive stars
exploding as Type II supernovae. Many studies of the Galactic disk stars 
support a view that [O/Fe] declines from [O/Fe]~$\sim$~0.5 at [Fe/H]~=~$-$1 
to about 0 at [Fe/H]~=~0 (Chen et al. \cite{Che00}; Edvardsson et al. \cite{Edv93}; 
Feltzing \& Gustafsson \cite{Fel98}). It is commonly accepted that 
this decline is due to the enhanced contribution of iron from Type Ia supernovae
at about [Fe/H]~=~$-$1. Similar trends (but not as steep as for oxygen)  
were found for other $\alpha$-elements.
It was also found that the differences in [$\alpha$/Fe] in the metallicity
range $-$0.8~$<$~[Fe/H]~$<$~$-$0.4 are correlated with the mean orbital 
galactocentric distance. Namely, stars with large Galactic orbits (exceeding
9\,kpc), have [$\alpha$/Fe] smaller than stars in the inner orbits. 

The situation is more complicated at metallicities [Fe/H]~$>$~$-$0.1.
Galactic viscous disk models of Tsujimoto et al. (\cite{Tsu95}) predict
[O/Fe] vs. [Fe/H] to flatten out at solar metallicities. 
However, observations of Feltzing \& Gustafsson (\cite{Fel98}) show that
the [O/Fe] decline continues even at [Fe/H]~$>$~0. It is interesting to
plot stars with planets on the same [O/Fe] vs [Fe/H] graph together
with the field metal rich stars in order to find out whether or not 
our targets follow the Galactic trend discussed by Feltzing \& Gustafsson.
Combining the [O/Fe] ratios reported in this paper together with the
values presented in the literature we found a large scatter on the
[O/Fe] vs [Fe/H] diagram (Fig.~\ref{figo}) for stars with [Fe/H]~$>$~0. It is 
interesting that \object{{\footnotesize HD}\,75289} and
\object{{\footnotesize HD}\,82943} have very similar abundances of all 
elements except oxygen which provides [O/Fe]~=~$-$0.30 and $-$0.07, respectively. 
Another interesting case was observed in \object{{\footnotesize HD}\,217107} 
where $\alpha$-elements Mg, Si and S are enhanced with respect to [Fe/H] by about 
0.2~dex while oxygen is again under-abundant (Sadakane et al. \cite{Sad99}). 
Note that for \object{{\footnotesize HD}\,83443} 
we derived [O/Fe]~=~0.12. The small number of observations and a use of only one oxygen 
line prevents us making from any firm conclusions. 
It would help to study different oxygen lines 
(IR triplets at 7775 and 8446\,\AA, OH bands in the near-UV and IR) in
order to obtain consistent abundances from the lines formed in different 
atmospheric layers.

\begin{figure}
\psfig{width=\hsize,file=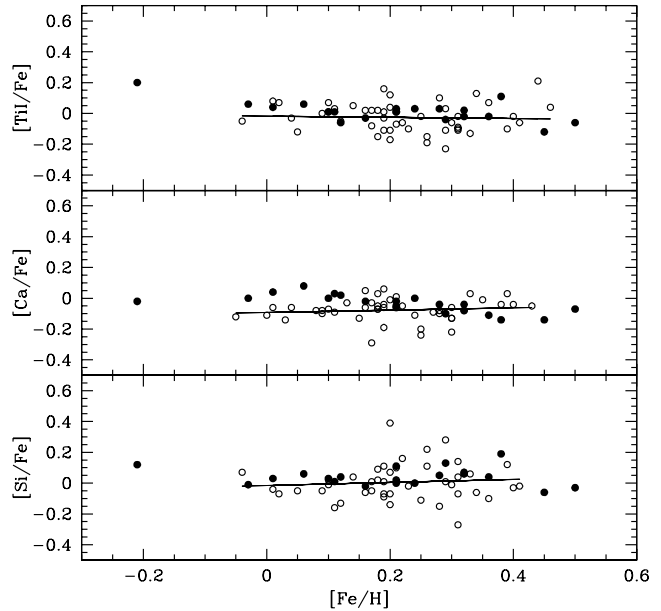}
\caption[]{Plots of [X/Fe] vs. [Fe/H] for three $\alpha$-elements. Open circles represent 
field stars from the study of Feltzing \& Gustafsson (\cite{Fel98}), and filled circles
stars with planetary mass companions. Fits to the field star sample are also shown.}
\label{figf}
\end{figure}

In Fig.~\ref{figf} we present the plots for [X/Fe] as a function of [Fe/H] for 
X~=~Ca, \ion{Ti}{i}, and Si, as compared with the values obtained by Feltzing \& Gustafsson (\cite{Fel98})
in their survey of 47 G and K dwarfs. The values of [X/H] for the stars with planets were 
taken from this paper and from Gonzalez \& Laws (\cite{Gon00}),  Gonzalez (\cite{Gon98b}), 
and Sadakane et al. (\cite{Sad99}).

An analysis of the figure shows that there are no apparent trends 
distinguishing stars with planets from single stars. This result does not exclude that
stars with planets may have a different ``behavior'' in such plots, but
it definitely shows that if they exist, then they must be of the order of 0.1~dex or lower
(considering the errors in the determinations of the abundances). 

Moreover, comparison of Ca and Ti abundances in the stars with planets with those 
from the field (Edvardsson et al. \cite{Edv93}; Feltzing \& Gustafsson \cite{Fel98})
confirms previous findings that [Ca/Fe] and [Ti/Fe] flattens out towards higher
metallicities. We confirm a small scatter in [Ca/Fe] and [Ti/Fe] found
by these authors. Our data for Si does not show a large scatter as found by 
Feltzing \& Gustafsson (\cite{Fel98}). Observations of Chen et al. (\cite{Che00}) support he small scatter found for Si in our study.

\subsection{Primordial abundance vs. Enrichment}

Various formation models try to take into account the observed anomalies 
to explain how and why the observed systems were formed and evolved.
In the continuation of the conventional picture where an ``ice'' core
is needed to accrete gas and give origin to a giant planet, the ``new'' theories 
include inward migration of the formed planet due 
to gravitational interaction with the disk (Goldreich \& Tremaine \cite{Gol80}; 
Lin \& Papaloizou \cite{Lin86}; Lissauer \cite{Lis95}), gravitational 
interactions between multiple giant planets (Weidenshilling \& Marzari \cite{Wei96}; 
Rasio \& Ford \cite{Ras96}) 
or even {\it in-situ} formation (e.g., Wuchterl \cite{Wuc96}; Bodenheimer et al. \cite{Bod99}), 
which was not compatible with former theories of giant gaseous planet formation. 
In these scenarios, the explanation of the higher metallicity found 
in giant extra-solar planet mother stars may involve mechanisms like 
the transfer of material from the disk to the star as the result of the migration processes (Lin \& Papaloizou \cite{Lin86}), or the fall of one or more
planets into the star. On the other hand, these anomalies can also be ``explained'' if 
we invoke that the formation of giant planets is dependent on the metallicity of the original 
molecular cloud. 

Since the mass of the convective envelope of a solar type dwarf increases with
increasing spectral type, if the enrichment scenario is the key of the observed 
chemical anomalies, we might expect that the value of [Fe/H] would be anti-correlated 
with the mass of the convective envelope at the time of planetary 
formation. This is particularly true for stars with planets in close orbits, since the migration 
process is expected to induce the fall of H and He poor disk material that was
inside the orbit of the planet (Lin et al. \cite{Lin96}).

\begin{figure}
\psfig{width=\hsize,file=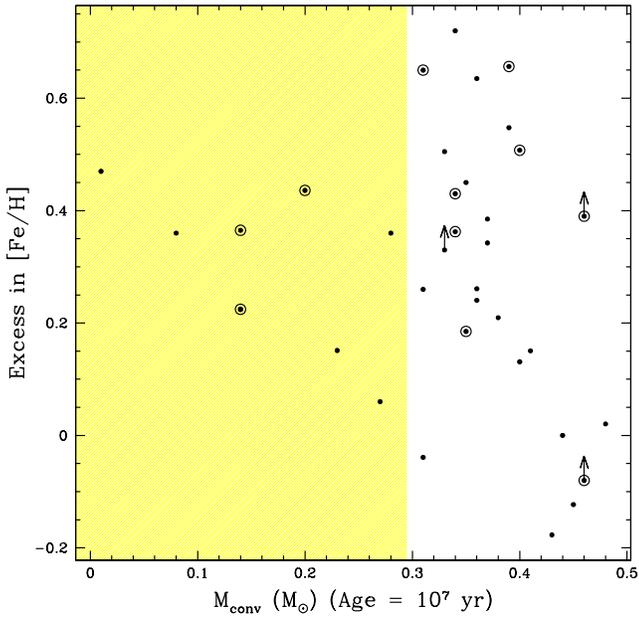}
\caption[]{Plot of the mass of the convective envelope as a function of [Fe/H]. ``Solar'' symbols denote
stars with planets orbiting closer than $\sim$0.08~AU. The shaded region represents the part of the diagram for which we believe there are observational biases. Arrows represent objects for which no age
was estimated, and an age of 0.0~yrs was considered.}
\label{fig4}
\end{figure}

\begin{figure}
\psfig{width=\hsize,file=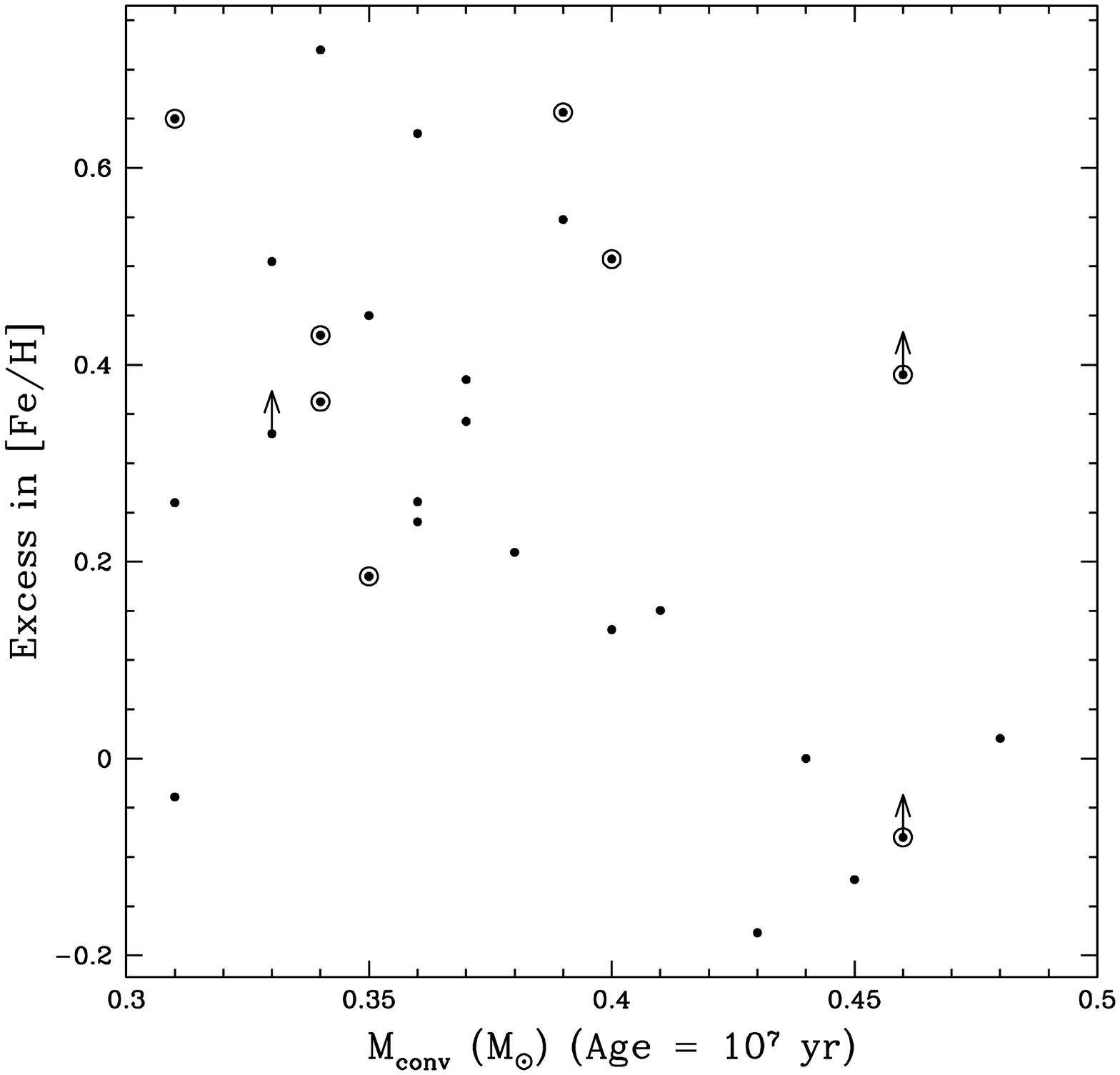}
\caption[]{Same as in Fig~\ref{fig4} but for the region with $M_\mathrm{conv}~\ge~$0.3~$M_{\sun}$.}
\label{fig5}
\end{figure}

To test this hypothesis we plot in Fig.~\ref{fig4} the values of the [Fe/H] excess
for the extra-solar planet harboring stars
listed in Table~\ref{tab6}, against the mass of the convection zone ($M_\mathrm{conv}$). 
The [Fe/H] excess is defined here as being the difference between the observed value and the
one expected according to the relation obtained by Gonzalez (\cite{Gon99}):
\begin{eqnarray}
\mathrm{[Fe/H]} & = & -0.035~\mathrm{Age(Gyr)} - 0.01
\label{eqn1}
\end{eqnarray}
This correction is done in order to take into account the galactic age gradient; it 
did not, however, change particularly the distribution of the stars in the diagram.
We do not make any correction for galactocentric distance. This does not, in principle,
introduce any systematic errors. 

The ages were taken from different authors (see Table~\ref{tab6}), and when 
no values were available, they were computed from the \ion{Ca}{ii} emission measure 
$R'_\mathrm{HK}$ using the relation of Donahue (\cite{Don93}), also quoted in Henry et 
al. (\cite{Hen96}). For three of the objects we have no value for the age: because of their position 
in the H-R diagram the errors computed using evolutionary tracks are very high; no 
$R'_\mathrm{HK}$ was available for these objects.

\begin{table*}
\caption[]{Values used to make the plots of Figures~\ref{fig4}, \ref{fig5}, and \ref{fig6}. For the references, G99, F99, and M00 correspond to Gonzalez (\cite{Gon99}), Fischer et al. (\cite{Fis99}), and Mazeh et al. (\cite{Maz00}), respectively.}
\begin{tabular}{lrlcccccc}
\hline
\noalign{\smallskip}
Star  & [Fe/H] & Age  & Age  & [Fe/H] & [Fe/H] & $M_{star}$ & $M_{c}$\,(10$^7$\,yr) & $M_{c}$\,(10$^8$\,yr) \\
      &  observed  & (Gyr)& Reference & from age &  excess &   (M$_{\sun}$)& (M$_{\sun}$) & (M$_{\sun}$) \\
\hline \\
\object{{\footnotesize BD}\,$-$10\,3166} &0.50 &4 &\ion{Ca}{ii} &-0.15 &0.65 &1.10 &0.31 &0.012\\
\object{{\footnotesize HD}\,1237} &0.10 &0.6 &\ion{Ca}{ii} &-0.03 &0.13 &0.96 &0.40 &0.037\\
\object{{\footnotesize HD}\,9826} &0.12 &2.7 &G99 &-0.10 &0.22 &1.30 &0.14 &0.000\\
\object{{\footnotesize HD}\,12661} &0.32 &-- &-- &-0.01 &0.33 &1.07 &0.33 &0.019\\
\object{{\footnotesize HD}\,13445} &-0.21 &2.2 &\ion{Ca}{ii} &-0.09 &-0.12 &0.86 &0.45 &0.050\\
\object{{\footnotesize HD}\,16141} &0.02 &6.6 &\ion{Ca}{ii} &-0.24 &0.26 &1.03 &0.36 &0.024\\
\object{{\footnotesize HD}\,17051} &0.11 &0.9 &\ion{Ca}{ii} &-0.04 &0.15 &1.19 &0.23 &0.000\\
\object{{\footnotesize HD}\,37124} &-0.32 &3.8 &\ion{Ca}{ii} &-0.14 &-0.18 &0.91 &0.43 &0.046\\
\object{{\footnotesize HD}\,46375} &0.34 &4.5 &\ion{Ca}{ii} &-0.17 &0.51 &0.96 &0.40 &0.037\\
\object{{\footnotesize HD}\,52265} &0.21 &4 &\ion{Ca}{ii} &-0.15 &0.36 &1.13 &0.28 &0.007\\
\object{{\footnotesize HD}\,75289} &0.23 &5.6 &\ion{Ca}{ii} &-0.21 &0.44 &1.22 &0.20 &0.000\\
\object{{\footnotesize HD}\,75732} &0.45 &5.0 &G99 &-0.19 &0.64 &1.03 &0.36 &0.023\\
\object{{\footnotesize HD}\,82943} &0.32 &5 &\ion{Ca}{ii} &-0.19 &0.51 &1.08 &0.40 &0.015\\
\object{{\footnotesize HD}\,83443} &0.38 &-- &-- &-0.01 &0.39 &0.85 &0.46 &0.051\\
\object{{\footnotesize HD}\,89744} &0.18 &8 &\ion{Ca}{ii} &-0.29 &0.47 &1.43 &0.01 &0.000\\
\object{{\footnotesize HD}\,95128} &0.01 &6.3 &G99 &-0.23 &0.24 &1.03 &0.36 &0.023\\
\object{{\footnotesize HD}\,108147} &-0.02 &2 &\ion{Ca}{ii} &-0.08 &0.06 &1.15 &0.31 &0.005\\
\object{{\footnotesize HD}\,117176} &-0.03 &8 &G99 &-0.29 &0.26 &1.10 &0.31 &0.012\\
\object{{\footnotesize HD}\,120136} &0.32 &1 &G99 &-0.05 &0.37 &1.30 &0.14 &0.000\\
\object{{\footnotesize HD}\,130322} &-0.02 &0.3 &\ion{Ca}{ii} &-0.02 &-0.00 &0.89 &0.44 &0.050\\
\object{{\footnotesize HD}\,134987} &0.23 &6.0 &\ion{Ca}{ii} &-0.22 &0.45 &1.05 &0.35 &0.021\\
\object{{\footnotesize HD}\,143761} &-0.29 &12.3 &G99 &-0.44 &0.15 &0.95 &0.41 &0.039\\
\object{{\footnotesize HD}\,145675} &0.50 &6 &G99 &-0.22 &0.72 &1.06 &0.34 &0.019\\
\object{{\footnotesize HD}\,168443} &-0.14 &2.6 &\ion{Ca}{ii} &-0.10 &-0.04 &1.10 &0.31 &0.012\\
\object{{\footnotesize HD}\,168746} &-0.09 &-- &-- &-0.01 &-0.08 &0.86 &0.41 &0.051\\
\object{{\footnotesize HD}\,169830} &0.21 &4 &\ion{Ca}{ii} &-0.15 &0.36 &1.37 &0.04 &0.000\\
\object{{\footnotesize HD}\,186427} &0.06 &9 &G99 &-0.33 &0.39 &1.01 &0.37 &0.027\\
\object{{\footnotesize HD}\,187123} &0.16 &5.5 &G99 &-0.20 &0.36 &1.06 &0.34 &0.019\\
\object{{\footnotesize HD}\,192263} &0.00 &0.3 &\ion{Ca}{ii} &-0.02 &0.02 &0.79 &0.48 &0.062\\
\object{{\footnotesize HD}\,195019} &0.00 &9.5 &F99 &-0.34 &0.34 &1.02 &0.37 &0.026\\
\object{{\footnotesize HD}\,209458} &0.00 &5 &M00 &-0.19 &0.19 &1.05 &0.35 &0.021\\
\object{{\footnotesize HD}\,210277} &0.24 &8.5 &G99 &-0.31 &0.55 &0.99 &0.39 &0.031\\
\object{{\footnotesize HD}\,217014} &0.21 &6 &G99 &-0.22 &0.43 &1.06 &0.34 &0.019\\
\object{{\footnotesize HD}\,217107} &0.30 &9.9 &F99 &-0.36 &0.66 &0.98 &0.39 &0.033\\
\object{{\footnotesize HD}\,222582} &0.00 &5.7 &\ion{Ca}{ii} &-0.21 &0.21 &1.00 &0.38 &0.029\\
\\
\noalign{\smallskip}
\hline
\end{tabular}
\label{tab6}
\end{table*}

The masses for the convective envelopes ($M_\mathrm{conv}$) were derived from Table~1 of D'Antona 
\& Mazzitelli (\cite{Dan94}), considering an age of 10$^7$~yr, and 
the stellar masses published by Butler et al. (\cite{But00}).
For the stars not listed by these authors,
stellar masses were derived from the position of the star in the evolutionary tracks of 
Schaller et al. (\cite{Sch92}). The age of 10$^7$~Gyr was taken to be the 
life time of a proto-planetary disk (Zuckerman et al. \cite{Zuc95}), and thus
a probable value for disk contamination to occur. 

One interesting feature to note in the plot is that the region with $M_\mathrm{conv}~\le~0.3$\,M$_{\sun}$
(shaded region) has a very low number of stars. But rather than having a physical origin,
we believe that the reason for this effect has to do with sampling effects. The stars
in this region are late-F and early-G dwarfs. Dwarfs of these spectral types are
usually fast rotators, consequently having higher intrinsic radial-velocity ``jitter''
(Mayor et al. \cite{May98}; Saar et al. \cite{Saa98}; Santos et al. \cite{San00}), and thus 
more difficult targets for high-precision radial-velocity searches for planets. 
Also, given their masses, a random sample of F dwarfs must be in general younger and thus more metal-rich than an equivalent sample of G dwarfs. This may explain the fact that 
stars in this region are slightly above the mean
[Fe/H] in the plot. Given these biases, we will concentrate on the right side of the diagram, 
plotted in more detail in Fig~\ref{fig5}.

This plot gives the impression that some trend exists in the sense of an anti-correlation. 
This is due to the presence of a very few stars in the lower-right corner of the diagram.
However, a linear fit to the points shows no significant correlation (we obtain a Spearman 
correlation coefficient of $-$0.3).  

It is important at this point to discuss the sources of uncertainties in the diagram.
First of all, the mass of the convection zone changes very fast with age in those 
evolutionary ages. For a 1\,M$_{\sun}$ star, if instead of 10$^7$ we take 3\,10$^7$\,yr,
the difference in the mass of the convection zone increases by about 0.3\,M$_{\sun}$.
It is possible that disk contamination can happen in slightly different time
scales for different stars, and thus the final result would be completely different.
Furthermore, the contamination scenario itself is not simple. If instead of disk contamination
we imagine that one or more planets fell into the star (by dynamical interactions
in a multiple system), this would not necessarily take place at the same time in a star's 
life. And of course, why should one expect that all stars would be ``polluted'' by the same amount of material? These facts may probably explain, or at least contribute, to the observed dispersion. 

Adding to these errors, the precision of eqn.~\ref{eqn1} is difficult to establish, but errors of
the order of 0.2 to 0.3 dex are expected. The errors in the age are also not measurable, but
might amount to some Gyr. To these we must add the uncertainties in the stellar models.

On the other hand, the analysis of the plot of Fig.~\ref{fig5} poses another problem: how can we explain that ``contamination'' effects may have increased the [Fe/H] by more than 0.6~dex in
stars with convection envelopes having masses around 0.35~$M_{\sun}$? For example,
the fall of 10 earth masses of iron into a star with a convective envelope
having solar metallicity and a mass of 0.3~$M_{\sun}$ would increase the [Fe/H]
by an insignificant amount. If the abundances are really due to some enrichment
process, then they must probably have taken place much after 10$^7$~yr. If the disk time-scales are correct, this would imply 
that the process did not involve the fall of disk material, but the addition of formed planets.

\begin{figure}
\psfig{width=\hsize,file=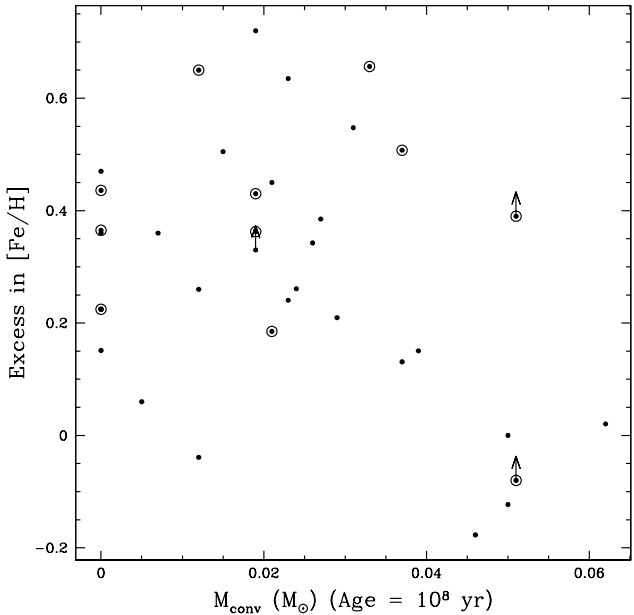}
\caption[]{Same as in Fig~\ref{fig4} but for $M_{conv}$ values at age = 10$^8$\,yr.}
\label{fig6}
\end{figure}

In Fig.~\ref{fig6} we have thus made a plot of the same variables, but this 
time $M_\mathrm{conv}$ was computed for an age of 10$^8$ yr. Still no 
correlation is evident. Although 
the size of the convection envelope has dropped, it seems unlikely that the high values observed for 
the [Fe/H] excess can be explained in an enrichment scenario (due to the fall of one or 
multiple planets). The convective outer envelope of the Sun at an age of 10$^8$~yr represents 
about 3\% of the total mass, and the fall of the same quantity of iron would change
the abundance by about $\sim$0.15~dex. This value would drop if we take a giant gaseous planet,
containing a certain amount of H and He. To have an excess of [Fe/H] of the order of 0.6~dex
we have to imagine that a star like the Sun had to swallow about 30 earth masses of iron.
Considering the composition of C1 chondrites, this would mean $\sim$5 times more in silicate material,
a value that seems excessively high. Unless multiple silicate-rich giant planets fell into the
star, this result clearly supports the idea 
that the cause of the excess of iron is probably ``primordial''. Since 
the high [Fe/H] vs. existence of planet relation is not in cause, these arguments favor a scenario where the formation of giant planets is dependent on the metallicity of the parent cloud.

\section{Concluding remarks}

We have presented an abundance analysis of 7 stars known to harbor giant planets and one with a 
brown dwarf candidate. The results obtained further support the idea that stars with planets are
metal rich when compared with field dwarfs. 

We discuss two possible scenarios that could explain the observed anomalies.
The results are still not conclusive, but support the idea that a star needs to be 
formed out of a metal-rich cloud to form giant planets.  

One other important conclusion comes out of the present work: to be able to
search and study hypothetical chemical anomalies in elements other than iron one must
lower the errors in the abundance determination to at least 0.1~dex.
This should be possible in the context of a high resolution and S/N spectroscopic study, 
using for all objects the same spectral lines, atmosphere models, and if possible making use 
of a comparison set of stars without planetary companions and with similar atmospheric 
parameters.

\begin{acknowledgements}
  We wish to thank the Swiss National Science Foundation (FNSRS) for
  the continuous support to this project. 
  We would also like to thank Claudio Melo, Daniel Ersparmer, Pierre North,
  Andr\'e Maeder and Bernard Pernier for useful discussions.
  Support from Funda\c{c}\~ao para a Ci\^encia e Tecnologia, Portugal, 
  to N.C.S. in the form of a scholarship is gratefully acknowledged.
\end{acknowledgements}


\end{document}